\documentclass[12pt,cite,epsf,epsfig]{article}
\usepackage{epsfig}

\begin{document}

\author{S. Dev\thanks{dev5703@yahoo.com}, Shivani Gupta\thanks{shiroberts\_1980@yahoo.co.in} and Radha Raman Gautam\thanks{gautamrrg@gmail.com}}

\title{Parametrizing the Lepton Mixing Matrix in terms of Charged Lepton Corrections}
\date{\textit{Department of Physics, Himachal Pradesh University, Shimla 171005, India.}\\
\smallskip}

\maketitle
\begin{abstract}
We consider a parametrization of the lepton mixing matrix in which the deviations from maximal atmospheric mixing and vanishing reactor mixing are obtained in terms of small corrections from the charged lepton sector. Relatively large deviations for the reactor mixing angle from zero as indicated by T2K experiment can be obtained in this parametrization. We are able to further reduce the number of complex phases, thus, simplifying the analysis. In addition, we have obtained the sides of unitarity triangles and the vacuum oscillation probabilities in this parametrization. The Jarlskog rephasing invariant measure of CP violation at the leading order has a single phase difference which can be identified as Dirac-type CP violating phase in this parametrization. 
\end{abstract}

\section{Introduction}
Results from a variety of solar, atmospheric and terrestrial neutrino oscillation experiments \cite{1} have constrained the form of the lepton mixing matrix $U$ \cite{2}. The lepton mixing matrix is given by
\begin{equation}
U = U_l^{\dagger} U_{\nu}
\end{equation}
where $U_l$ and $U_{\nu}$ are both 3$\times$3 unitary matrices such that $U_l$ arises from the diagonalization of the charged lepton mass matrix $(M_l)$ while $U_\nu$ diagonalizes the neutrino mass matrix $(M_\nu)$. For three lepton generations, the 3$\times$3 unitary matrix $U$ in Particle Data Group (PDG) \cite{3} parametrization is given by
\begin{equation}
U = \left(
\begin{array}{ccc}
c_{12}c_{13} & s_{12}c_{13} & s_{13}e^{-i\delta} \\
-s_{12}c_{23}-c_{12}s_{23}s_{13}e^{i\delta} &
c_{12}c_{23}-s_{12}s_{23}s_{13}e^{i\delta} & s_{23}c_{13} \\
s_{12}s_{23}-c_{12}c_{23}s_{13}e^{i\delta} &
-c_{12}s_{23}-s_{12}c_{23}s_{13}e^{i\delta} & c_{23}c_{13}
\end{array}
\right).\wp
\end{equation}
where $s_{13}$ = $\sin \theta_{13}$, $c_{13}$ = $\cos \theta_{13}$ with $\theta_{13}$ being the reactor angle, $s_{12}$ = $\sin \theta_{12}$, $c_{12}$ = $\cos \theta_{12}$ with $\theta_{12}$ being the solar angle, $s_{23}$ = $\sin \theta_{23}$, $c_{23}$ = $\cos \theta_{23}$ with $\theta_{23}$ being the atmospheric mixing angle and $\delta$ is the Dirac-type CP violating phase. The phase matrix $\wp$ = diag$(1,e^{i \alpha_1/2},e^{i \alpha_2/2})$ contains the Majorana-type CP violating phases $\alpha_1$ and $\alpha_2$ which do not affect neutrino oscillations and are not directly accessible to experimental scrutiny at present. Current data is consistent with the tribimaximal (TBM) mixing \cite{4} 
\begin{equation}
U_{TBM} = \left(
\begin{array}{ccc}
\frac{2}{\sqrt{6}} & \frac{1}{\sqrt{3}} & 0 \\
\frac{-1}{\sqrt{6}}&
\frac{1}{\sqrt{3}}& \frac{-1}{\sqrt{2}}\\
\frac{-1}{\sqrt{6}}&
\frac{1}{\sqrt{3}}& \frac{1}{\sqrt{2}}\\
\end{array}
\right).\wp
\end{equation} 
which has been derived using family symmetries \cite{5}. In addition to TBM, there are other mixing schemes which can reproduce the observed leptonic mixing pattern which include the two Golden Ratio (GR) mixing schemes where the mixing angles are for GR1: $\theta_{12}$ = $\tan^{-1}(1/\varphi)$, $\theta_{23}$ = $\pi/4$, $\theta_{13}$ = 0 \cite{6}, GR2: $\theta_{12}$ = $\cos^{-1}(\varphi/2)$, $\theta_{23}$ = $\pi/4$, $\theta_{13}$ = 0 \cite{7} where $\varphi$ = $(1+\sqrt{5})/2$, Hexagonal Mixing (HM): $\theta_{12}$ = $\pi/6$, $\theta_{23}$ = $\pi/4$, $\theta_{13}$ = 0 \cite{8}, Bimaximal Mixing (BM): $\theta_{12}$ = $\pi/4$, $\theta_{23}$ = $\pi/4$, $\theta_{13}$ = 0 \cite{9}. All these mixing schemes can arise from mass independent textures also known as form diagonalizable textures \cite{10} and lead to a predictive neutrino mass matrix structure which contains just five parameters (the three neutrino masses and two Majorana phases). All the above mixing scenarios have the same predictions for the reactor and atmospheric mixing angles viz  $\theta_{13}$ = 0 and $\theta_{23}$ = $\pi/4$, whereas their predictions for the solar mixing angle $\theta_{12}$ are different. Thus, the above mixing matrices are common upto a mixing matrix
\begin{equation}
U = \left(
\begin{array}{ccc}
c_{12}' & s_{12}' & 0 \\
\frac{-s_{12}'}{\sqrt{2}}&
\frac{c_{12}'}{\sqrt{2}}& \frac{-1}{\sqrt{2}}\\
\frac{-s_{12}'}{\sqrt{2}}&
\frac{c_{12}'}{\sqrt{2}}& \frac{1}{\sqrt{2}}\\
\end{array}
\right).\wp
\end{equation}   
arising from a mu-tau symmetric neutrino mass matrix. It is highly unlikely that any of the above mixing schemes is exact since there are already hints of a non-zero reactor mixing angle $\theta_{13}$ \cite{11}. Recently, the T2K collaboration has observed possible indications of the $\nu_{\mu}\rightarrow\nu_{e}$ appearance and reported the following ranges for $\theta_{13}$ \cite{12}
\begin{equation}
5.0^o<\theta_{13}<16.0^o
for NH \end{equation} 
\begin{equation}
5.8^o<\theta_{13}<17.8^o
for IH \end{equation} at $90\%$ C.L.. Moreover, the best fit value of $\theta_{13}$ is found to be $\theta_{13}\approx9.7^o$ for NH and $\theta_{13}\approx11^o$ of IH, thus, implying large deviations from $\theta_{13}=0^o$ in the above mentioned mixing scenarios.
Therefore, it becomes important to develop a parametrization of the lepton mixing matrix in which such deviations are manifest. A natural possibility to obtain a phenomenologically viable neutrino mixing matrix and to generate non-zero $\theta_{13}$ and non-maximal $\theta_{23}$ is to assume that these deviations come from the charged lepton sector. Such an assumption has been made earlier to generate deviations from bimaximal mixing \cite{13,14} and tribimaximal mixing \cite{15,16,17,18}. 

\section{Formalism}
A general 3$\times$3 matrix contains 3 moduli and 6 phases \cite{19} and can be represented as
\begin{equation}
U = e^{i \Phi} P \tilde{U} Q
\end{equation}
where $P$ = diag$(1,e^{i \phi_1},e^{i \phi_2})$ and $Q$ = diag$(1,e^{i \rho_1},e^{i \rho_2})$ are diagonal phase matrices having two phases each, $\tilde{U}$ is the matrix containing 3 angles and one phase and has the form of $U$ (except for the phase matrix $\wp$) in Eq.(2). In general, when charged leptons also contribute to the mixing, the lepton mixing matrix contains 6 real parameters and six phases \cite{14}. As pointed out earlier the two Majorana phases are unlikely to be measured in the present and the forthcoming experiments, so that we may dispense with the Majorana phases at least for the present by considering the Hermitian products $M_l M_l^{\dagger}$ and $M_\nu M_\nu^{\dagger}$. Here, two points are in order:\\
1.) Since we are considering mass independent textures of the neutrino mass matrix, thus, $M_\nu$ and $M_\nu M_\nu^{\dagger}$ are diagonalized by the same diagonalizing matrix, so we can consider the product $M_\nu M_\nu^{\dagger}$.\\
2.) The deviations of the charged lepton mass matrix from diagonal matrix are in any case considered to be arbitrary so the choice of product $M_l M_l^{\dagger}$ can be made.\\ By using these Hermitian products we not only dispense with the unnecessary burden of Majorana phases but we are also able to remove one additional phase from the lepton mixing matrix, thus, simplifying the subsequent analysis. The lepton mass matrices can be diagonalized as
\begin{equation}
M_l = U_l M_l^d U_R^{\dagger}
\end{equation}
\begin{equation}
M_\nu = U_\nu M_\nu^d U_\nu ^T.
\end{equation}
Thus, the product $M_l M_l^{\dagger}$ becomes
\begin{eqnarray}
M_l M_l^{\dagger} = U_l M_l^d U_R^{\dagger} U_R M_l^d U_l^{\dagger}\\ \nonumber
= U_l (M_l^d)^2 U_l^{\dagger} 
\end{eqnarray}
which can be written as
\begin{eqnarray}
M_l M_l^{\dagger} = e^{i \phi_l} P_l \tilde{U_l} Q_l (M_l^d)^2 Q_l^{\dagger} \tilde{U_l}^{\dagger} P_l^{\dagger} e^{-i \phi_l}\\ \nonumber
=P_l \tilde{U_l} (M_l^d)^2 \tilde{U_l}^{\dagger} P_l^{\dagger}
\end{eqnarray}
using Eq.(7). Similarly, for the product $M_\nu M_\nu^{\dagger}$ we obtain
\begin{eqnarray}
M_\nu M_\nu ^{\dagger} = P_\nu \tilde{U_\nu} (M_\nu ^d)^2 \tilde{U_\nu}^{\dagger} P_\nu ^{\dagger}.
\end{eqnarray} 
We can absorb two phases from $P_l$ and one phase from $P_\nu$ in the left handed lepton fields and the resulting lepton mixing matrix is given by
\begin{equation}
U = \tilde{U_l}^{\dagger} P_\nu \tilde{U_\nu}
\end{equation}
where $\tilde{U_l}$ and $\tilde{U_\nu}$ contain three real parameters and one phase each while $P_\nu$ contains one phase $P_\nu$ = diag$(1,1,e^{i \phi})$. Thus, in this formalism $U$ is expressed in terms of six real parameters and three phases.\\
In the present work, we discuss a parametrization of the lepton mixing matrix which allows for the large deviations from $\theta_{13}=0$ and has the form of Eq.(4) at zeroth order. Here $\theta_{12}'$ can have the values $\sin^{-1}(1/\sqrt{3})$ for TBM mixing, $\tan^{-1}(1/\varphi)$ for GR1 mixing, $\cos^{-1}(\varphi/2)$ for GR2 mixing where $\varphi$ = $(1+\sqrt{5})/2$, $\pi/6$ for hexagonal mixing and $\pi/4$ for bimaximal mixing. Deviations from the above mentioned scenarios are parametrized in terms of charged lepton corrections represented by small parameters having magnitude of the order of Wolfenstein parameter $\lambda \approx 0.227$ or less. In the small angle approximation, we have to the first order in $\epsilon _{ij}$
\begin{equation}
\sin\epsilon _{ij} \approx \epsilon _{ij}, \ \cos\epsilon _{ij} \approx 1  \ \ \  \ \epsilon _{ij} < 0.227
\end{equation}
where $i, j$ = 1,2,3 and $i<j$.

\section{Deviations from exact mixing schemes}
In this section, we obtain expressions for lepton mixing observables in terms of the charged lepton corrections. The neutrino mixing matrix $\tilde{U_{\nu}}$ is assumed to have the form of Eq.(4) except for the phase matrix $\wp$. The charged lepton mixing matrix to first order \footnote{second order corrections to the mixing matrix elements and other relevant quantities have been discussed in the Appendix } in terms of small parameters is given by
\begin{equation}
\tilde{U_l} = \left(
\begin{array}{ccc}
 1 & \epsilon _{12} & e^{-i \delta _{13}}\epsilon _{13} \\
 -\epsilon _{12}& 1 & \epsilon _{23} \\
 -e^{i \delta _{13}} \epsilon _{13} & -\epsilon _{23} & 1
\end{array}
\right)
\end{equation} 
and the resulting lepton mixing matrix has the form 
\begin{eqnarray} 
U = \tilde{U_l}^{\dagger} P_\nu \tilde{U_\nu} = \nonumber
\left(
\begin{array}{ccc}
c_{12}' & s_{12}' & 0 \\
\frac{-s_{12}'}{\sqrt{2}}&
\frac{c_{12}'}{\sqrt{2}}& \frac{-1}{\sqrt{2}}\\
\frac{- e^{i \phi}s_{12}'}{\sqrt{2}}&
\frac{ e^{i \phi}c_{12}'}{\sqrt{2}}& \frac{e^{i \phi}}{\sqrt{2}}\\
\end{array}
\right) + \\ 
\left(
\begin{array}{ccc}
\frac{s_{12}'(\epsilon_{12} +e^{-i (\delta_{13} - \phi)} \epsilon_{13})}{\sqrt{2}} & - \frac{c_{12}' (\epsilon_{12} +e^{-i (\delta_{13} - \phi)} \epsilon_{13})}{\sqrt{2}} & \frac{\epsilon_{12} - e^{-i (\delta_{13} - \phi)} \epsilon_{13}}{\sqrt{2}}\\
c_{12}' \epsilon_{12} + \frac{e^{i \phi} s_{12}' \epsilon_{23}}{\sqrt{2}} & s_{12}' \epsilon_{12} - \frac{e^{i \phi} c_{12}' \epsilon_{23}}{\sqrt{2}} & \frac{- e^{i \phi}\epsilon_{23}}{\sqrt{2}} \\
\frac{-s_{12}' \epsilon_{23}}{\sqrt{2}} + e^{i \delta _{13}} c_{12}'\epsilon_{13} & \frac{c_{12}' \epsilon_{23}}{\sqrt{2}} + e^{i \delta_{13}} s_{12}'\epsilon_{13} & \frac{-\epsilon_{23}}{\sqrt{2}}
\end{array}
\right).
\end{eqnarray}
The lepton mixing angles are related to the elements of the mixing matrix as
\begin{eqnarray}
\sin^2\theta_{13} = |U_{e3}|^2, \ \ \sin^2\theta_{23} = \frac{|U_{\mu3}|^2}{|U_{\mu3}|^2+|U_{\tau3}|^2}, \  \ \sin^2\theta_{12} = \frac{|U_{e2}|^2}{|U_{e1}|^2+|U_{e2}|^2}.
\end{eqnarray}
The mixing angles in this parametrization, to first order in small parameters are given by
\begin{eqnarray}
\sin\theta_{13} = \frac{\epsilon_{13} - \epsilon_{12} \cos(\delta_{13}-\phi)}{\sqrt{2}}, \nonumber \\
\sin \theta_{23} = \frac{1 + \epsilon_{23} \cos\phi}{\sqrt{2}},  \\
\sin \theta_{12} = s_{12}' -\frac{c_{12}' (\epsilon_{12} + \epsilon_{13}\cos(\delta_{13}-\phi))}{\sqrt{2}}. \nonumber
\end{eqnarray}
It can be seen that the reactor mixing angle along with the atmospheric mixing angle is independent of $\theta_{12}'$ and the deviation of the atmospheric mixing angle from maximality depends only on the small parameter $\epsilon_{23}$ in the first order corrections. We restrict the ranges of perturbation parameters by using the recent global analysis \cite{20} which incorporates the T2K \cite{12} and MINOS \cite{21} results. Allowed numerical ranges for the perturbation parameters $\epsilon_{13}$ and $\epsilon_{23}$ at 3$\sigma$ for all mixing scenarios are $-0.22<(\epsilon_{13},\epsilon_{23})<0.22$. The range of $\epsilon_{12}$ at 3$\sigma$ is\\
\begin{eqnarray}
-0.20<\epsilon_{12}<0.17 \ \ TBM \nonumber\\
-0.165<\epsilon_{12}<0.22 \ \ GR1 \nonumber\\
-0.22<\epsilon_{12}<0.17 \ \ GR2 \nonumber\\
-0.15<\epsilon_{12}<0.22 \ \ HM \nonumber\\
-0.22<\epsilon_{12}<0 \ \ BM. 
\end{eqnarray}
The Jarskog CP violation rephasing invariant \cite{22} is given by
\begin{equation}
J_{CP} = \frac{\sin2\theta_{12}' \epsilon_{13} \sin(\delta_{13}-\phi)}{4\sqrt{2}}.
\end{equation}
An important point to note here is that the Jarskog rephasing invariant $J_{CP}$ to the first order contains a single phase difference which is the relevant Dirac-type CP violating phase in this case. However, it does not necessarily coincide with the Dirac-type phase in the standard parametrization as pointed out in Ref. \cite{17}. Also, in  Ref. \cite{16,17}, the expression for $J_{CP}$ contains two different phases in the leading order term and further assuming CKM like hierarchy among the perturbation parameters, the relevant Dirac phase in these works comes out to be different from that in our parametrization.\\
Now we discuss unitarity triangles and neutrino oscillation formulae which get simplified using this parametrization. The unitarity triangles may be constructed using the orthogonality of different pairs of columns or rows of the mixing matrix. Information about the elements $U_{e2}$, $U_{e3}$ and $U_{\mu3}$ is obtained in the solar, reactor and atmospheric experiments and the most important unitarity triangles should include all these elements \cite{23}. Two such unitarity triangles correspond to the orthogonality of second and third column $\nu_2.\nu_3$ and orthogonality of the first and second row $\nu_e.\nu_{\mu}$. The unitarity relation for the $\nu_2.\nu_3$ triangle is given by
\begin{equation}
U_{e2} U_{e3}^* + U_{\mu2} U_{\mu3}^* + U_{\tau2} U_{\tau3}^* = 0
\end{equation}
and the sides of this unitarity triangle to first order are given by
\begin{eqnarray}
U_{e2} U_{e3}^* \approx \frac{\epsilon_{12}-e^{i(\delta_{13}-\phi)}s_{12}'\epsilon_{12}}{\sqrt{2}}, \nonumber \\
U_{\mu2} U_{\mu3}^* \approx -\frac{c_{12}'}{2}-\frac{s_{12}' \epsilon_{12}}{\sqrt{2}}+ i c_{12}' \epsilon_{23} \sin\phi, \\
U_{\tau2} U_{\tau3}^* \approx \frac{c_{12}'}{2}+\frac{e^{i(\delta_{13}-\phi)}s_{12}' \epsilon_{13}}{\sqrt{2}} - i c_{12}' \epsilon_{23} \sin\phi, \nonumber 
\end{eqnarray}
which satisfies Eq.(21) to first order. The invariant $J_{CP}$ is
\begin{equation}
J_{CP}=Im(U_{e2} U_{e3}^* U_{\mu2}^* U_{\mu3})=Im(U_{\tau2} U_{\tau3}^* U_{e2}^* U_{e3})=Im(U_{\mu2} U_{\mu3}^* U_{\tau2}^* U_{\tau3}). 
\end{equation}\\
The other unitarity triangle $\nu_e .\nu_{\mu}$ corresponds to the unitarity relation
\begin{equation}
U_{\mu1} U_{e1}^* + U_{\mu2} U_{e2}^* + U_{\mu3} U_{e3}^* = 0
\end{equation}
and to first order the sides of this unitarity triangle are given by
\begin{eqnarray}
U_{\mu1} U_{e1}^* \approx \frac{\sin2\theta_{12}'(-1+e^{i\phi}\epsilon_{23})}{2\sqrt{2}}+\frac{\epsilon_{12}(1+3\cos2\theta_{12}')}{4}-\frac{e^{i(\delta_{13}-\phi)}s_{12}'^{2} \epsilon_{13}}{2}, \nonumber \\
U_{\mu2} U_{e2}^* \approx -\frac{\sin2\theta_{12}'(-1+e^{i\phi}\epsilon_{23})}{2\sqrt{2}}+\frac{\epsilon_{12}(1-3\cos2\theta_{12}')}{4}-\frac{e^{i(\delta_{13}-\phi)}c_{12}'^{2} \epsilon_{13}}{2}, \\
U_{\mu3} U_{e3}^* \approx -\frac{\epsilon_{12}}{2}+\frac{e^{i(\delta_{13}-\phi)}\epsilon_{13}}{2}  \  \   \   \  \nonumber 
\end{eqnarray}
which satisfy Eq.(24) to first order. The invariant $J_{CP}$ is
\begin{equation}
J_{CP}=Im(U_{\mu3} U_{e3}^* U_{\mu2}^* U_{e2})=Im(U_{\mu1} U_{e1}^* U_{\mu3}^* U_{e3})=Im(U_{\mu2} U_{e2}^* U_{\mu1}^* U_{e1}). 
\end{equation}
 Now, we discuss the applications of this parametrization to neutrino oscillations. The probability of oscillation from flavor $\nu_{\alpha}$ to flavor $\nu_{\beta}$, $P(\nu_{\alpha} \rightarrow \nu_{\beta})$ is given by
\begin{eqnarray}
P(\nu_{\alpha} \rightarrow \nu_{\beta}) = \mid \sum_{i=1}^3 U_{\alpha i}^* e^{-im_i^2 \frac{L}{2E}} U_{\beta i} \mid^2  \nonumber  \\
=\delta_{\alpha \beta} - 4 \sum_{i>j} Re(U_{\alpha i}^* U_{\beta i} U_{\alpha j} U_{\beta j}^*) \sin^2 \triangle_{ij} + 2 \sum_{i>j} Im(U_{\alpha i}^* U_{\beta i} U_{\alpha j} U_{\beta j}^*) \sin2 \triangle_{ij}
\end{eqnarray}
where $\alpha$, $\beta$ = $e$, $\mu$, $\tau$, $\triangle_{ij} \equiv (m_i^2 - m_j^2)L/4E$, $L$ is the oscillation length and $E$ is the beam energy of neutrinos. Expanding to second order in $\epsilon_{ij}$ and $\triangle_{21}$ assuming $\triangle_{21} \ll 1$ \cite{23}, we obtain the various vacuum oscillation probabilities.\\
For $e\rightarrow e,\mu,\tau$ we obtain
\begin{eqnarray}
P(\nu_{e} \rightarrow \nu_{e}) = 1 - \triangle^2_{21}\sin^22\theta_{12}' -2(\epsilon_{12}^2+ s_{12}'^{2} \epsilon_{13}^2)\sin^2\triangle_{31} + 4 \epsilon_{12}\epsilon_{13} \cos(\delta_{13}-\phi) \sin^2\triangle_{31}, \nonumber \\
P(\nu_{e} \rightarrow \nu_{\mu}) = \frac{\triangle_{21}^2 \sin^22\theta_{12}'}{2}+ (\epsilon_{12}^2+\epsilon_{13}^2)\sin^2\triangle_{31}-2 \epsilon_{12}\epsilon_{13}\cos(\delta_{13}-\phi)\sin^2\triangle_{31}+ \nonumber \\ \frac{\triangle_{21} \sin2\theta_{12}' \epsilon_{12}\sin(\delta_{13}-\phi)}{\sqrt{2}},  \nonumber \\
P(\nu_{e} \rightarrow \nu_{\tau}) = \frac{\triangle_{21}^2 \sin^22\theta_{12}'}{2}+ (\epsilon_{12}^2+\epsilon_{13}^2)\sin^2\triangle_{31}-2 \epsilon_{12}\epsilon_{13}\cos(\delta_{13}-\phi)\sin^2\triangle_{31} \nonumber \\ -\frac{\triangle_{21} \sin2\theta_{12}' \epsilon_{13}\sin(\delta_{13}-\phi)}{\sqrt{2}}.
\end{eqnarray}
The above equations give electron neutrino survival and disappearance probabilities to the second order in $\epsilon_{ij}$ and $\triangle_{21}$. Note that these probabilities are independent of deviations from atmospheric mixing so that any deviation from maximal atmospheric mixing only appears at third order in these oscillation probabilities. \\
For $\mu\rightarrow e,\mu,\tau$ we obtain
\begin{eqnarray}
P(\nu_{\mu} \rightarrow \nu_{e}) = \frac{\triangle_{21}^2 \sin^22\theta_{12}'}{2}+ (\epsilon_{12}^2+\epsilon_{13}^2)\sin^2\triangle_{31}-2 \epsilon_{12}\epsilon_{13}\cos(\delta_{13}-\phi)\sin^2\triangle_{31}\nonumber \\-\frac{\triangle_{21} \sin2\theta_{12}' \epsilon_{13}\sin(\delta_{13}-\phi)}{\sqrt{2}}, \nonumber \\
P(\nu_{\mu} \rightarrow \nu_{\mu}) = 1 - \sin^2 \triangle_{31} - \frac{\triangle_{21}^2 \sin^22\theta_{12}'}{4} + 4 \epsilon^2_{23} \cos^2\phi \sin^2 \triangle_{31}^2, \nonumber \\
P(\nu_{\mu} \rightarrow \nu_{\tau}) = \sin^2 \triangle_{31} - \frac{\triangle_{21}^2 \sin^22\theta_{12}'}{4} - 4 \epsilon^2_{23} \cos^2\phi \sin^2 \triangle_{31}^2 - (\epsilon_{12}^2 \nonumber \\ +\epsilon_{13}^2)\sin^2\triangle_{31}+ 2 \epsilon_{12}\epsilon_{13}\cos(\delta_{13}-\phi)\sin^2\triangle_{31} + \frac{\triangle_{21} \sin2\theta_{12}' \epsilon_{13}\sin(\delta_{13}-\phi)}{\sqrt{2}}.
\end{eqnarray}
New results have been announced by long baseline experiment T2K probing the $\nu_{\mu}\rightarrow \nu_e$ appearance channel giving non zero reactor mixing angle \cite{12}. The deviation from maximal atmospheric mixing does not appear upto second order in the oscillation probability $P(\nu_{\mu}\rightarrow \nu_e)$. Therefore, in this parametrization $\theta_{13}$ can have a large deviation from zero irrespective of the deviation of $\theta_{23}$ from $\frac{\pi}{4}$ to a good approximation. \\
For $\tau\rightarrow e,\mu,\tau$ we have
\begin{eqnarray}
P(\nu_{\tau} \rightarrow \nu_{e}) = \frac{\triangle_{21}^2 \sin^22\theta_{12}'}{2}+ (\epsilon_{12}^2+\epsilon_{13}^2)\sin^2\triangle_{31}-2 \epsilon_{12}\epsilon_{13}\cos(\delta_{13}-\phi)\sin^2\triangle_{31}+ \nonumber \\ \frac{\triangle_{21} \sin2\theta_{12}' \epsilon_{12}\sin(\delta_{13}-\phi)}{\sqrt{2}},  \nonumber \\
P(\nu_{\tau} \rightarrow \nu_{\mu}) = \sin^2 \triangle_{31} - \frac{\triangle_{21}^2 \sin^22\theta_{12}'}{4}  - 4 \epsilon^2_{23} \cos^2\phi \sin^2 \triangle_{31}^2 - (\epsilon_{12}^2+\epsilon_{13}^2)\sin^2\triangle_{31}  \nonumber \\ + 2 \epsilon_{12}\epsilon_{13}\cos(\delta_{13}-\phi)\sin^2\triangle_{31} - \frac{\triangle_{21} \sin2\theta_{12}' \epsilon_{13}\sin(\delta_{13}-\phi)}{\sqrt{2}},\nonumber \\
P(\nu_{\tau} \rightarrow \nu_{\tau}) = 1 - \sin^2 \triangle_{31} - \frac{\triangle_{21}^2 \sin^22\theta_{12}'}{4} + 4 \epsilon^2_{23} \cos^2\phi \sin^2 \triangle_{31}^2. 
\end{eqnarray}
According to the above equations, some oscillation probabilities are identical upto the second order. This is the consequence of the simplifying assumptions $\triangle_{32}\approx\triangle_{31}$. These oscillation probabilities will differ slightly when third order perturbations are considered. 
\section{Summary}
Neutrino oscillation experiments suggest an atmospheric mixing angle very close to $\pi/4$ and a small reactor mixing angle. We assume the lepton mixing matrix at zeroth order having $\theta_{23}=\frac{\pi}{4}$ and $\theta_{13}=0$. Deviations from maximal atmospheric mixing and vanishing reactor mixing are obtained through charged lepton corrections in terms of small perturbation parameters. Relatively large deviations for the reactor mixing angle from zero as indicated by T2K experiment can be obtained in this parametrization. In the zeroth order lepton mixing matrix, we keep the solar mixing angle general, so that the deviations from a particular mixing scheme e.g. TBM, GR1, GR2, HM and BM can be obtained by substituting the value of solar mixing angle.
In this analysis, we have been able to reduce the number of complex phases to two by considering the Hermitian products of charged lepton and neutrino mass matrices, thus, resulting in considerable simplification of our analysis. The Jarlskog rephasing invariant measure of CP violation contains a single phase difference in the leading order which allows us to identify this phase difference with the Dirac-type CP violating phase in this parametrization. We have also obtained the formulae for the sides of unitarity triangles and vacuum oscillation probabilities. It is found that the deviation from maximal atmospheric mixing does not appear upto second order in the oscillation probability $P(\nu_{\mu}\rightarrow \nu_e)$ relevant for the measurement of reactor mixing angle. Therefore, in this parametrization $\theta_{13}$ can have a large deviation from zero irrespective of the deviation of $\theta_{23}$ from maximality to a good approximation. 

\textbf{\textit{\Large{Acknowledgements}}}

The research work of S. D. is supported by the University Grants
Commission, Government of India \textit{vide} Grant No. 34-32/2008
(SR). R. R. G. acknowledge the financial support provided by the Council for Scientific and Industrial Research (CSIR), Government of India.
\appendix
\section{Appendix}
Here, we list the second order corrections to the results given in the main text.
The second order corrections to the first order mixing matrix elements are given by
\begin{eqnarray}
\triangle U_{e1} \approx -\frac{c_{12}'(\epsilon^2_{12}+\epsilon^2_{13})}{2}+\frac{s_{12}' \epsilon_{23}(e^{-i\delta_{13}} \epsilon_{13}-e^{i\phi} \epsilon_{12})}{\sqrt{2}}, \nonumber \\
\triangle U_{e2} \approx -\frac{s_{12}'(\epsilon^2_{12}+\epsilon^2_{13})}{2}+\frac{c_{12}' \epsilon_{23}(e^{i\phi} \epsilon_{12}-e^{-i\delta_{13}} \epsilon_{13})}{\sqrt{2}}, \nonumber \\
\triangle U_{e3} \approx \frac{e^{i\phi}\epsilon_{12}\epsilon_{23}+e^{-i\delta_{13}}\epsilon_{13}\epsilon_{23}}{\sqrt{2}}, \nonumber \\
\triangle U_{\mu1} \approx \frac{s_{12}'(\epsilon^2_{12}+\epsilon^2_{23})}{2\sqrt{2}}+\frac{e^{-i(\delta_{13}-\phi)}s_{12}' \epsilon_{12}\epsilon_{13}}{\sqrt{2}}, \nonumber \\
\triangle U_{\mu2} \approx \frac{c_{12}'(\epsilon^2_{12}+\epsilon^2_{23})}{2\sqrt{2}}-\frac{e^{-i(\delta_{13}-\phi)}c_{12}' \epsilon_{12}\epsilon_{13}}{\sqrt{2}}, \nonumber \\
\triangle U_{\mu3} \approx \frac{(\epsilon^2_{12}+\epsilon^2_{23})}{2\sqrt{2}}-\frac{e^{-i(\delta_{13}-\phi)} \epsilon_{12}\epsilon_{13}}{\sqrt{2}}, \nonumber \\
\triangle U_{\tau1} \approx \frac{e^{i\phi} s_{12}' (\epsilon^2_{13}+ \epsilon^2_{23})}{2\sqrt{2}}, \nonumber \\
\triangle U_{\tau2} \approx -\frac{e^{i\phi} c_{12}'(\epsilon^2_{13}+ \epsilon^2_{23})}{2\sqrt{2}}, \nonumber \\
\triangle U_{\tau3} \approx -\frac{e^{i\phi} (\epsilon^2_{13}+ \epsilon^2_{23})}{2\sqrt{2}}.  
\end{eqnarray}
The second order corrections to the mixing angles are given by
\begin{eqnarray}
\triangle \sin\theta_{13} \approx \frac{\epsilon_{23}\cos\phi (\epsilon_{13}+\epsilon_{12}\cos(\delta_{13}-\phi))}{\sqrt{2}} + \frac{\epsilon^2_{12}\sin^2(\delta_{13}-\phi)(1+3 \epsilon_{23}\cos\phi)}{\epsilon_{13}2\sqrt{2}}, \nonumber \\
\triangle \sin\theta_{23} \approx -\frac{\epsilon^2_{12}-\epsilon^2_{13}+\epsilon^2_{23}- 2\epsilon_{12}\epsilon_{13}\cos(\delta_{13}-\phi) + \epsilon^2_{23}\cos2\phi}{4\sqrt{2}}, \nonumber \\
\triangle \sin\theta_{12} \approx \frac{\epsilon^2_{13}(\csc \theta_{12}' - 3s_{12}' - c_{12}' \cot\theta_{12}' \cos2(\delta_{13}-\phi)) - 2s_{12}' \epsilon^2_{12}}{8} + \nonumber \\ \frac{\sqrt{2}c_{12}' \epsilon_{12}\epsilon_{23}\cos\phi - \epsilon_{13}(\sqrt{2}c_{12}' \epsilon_{23}\cos\delta_{13} + s_{12}' \epsilon_{12}\cos(\delta_{13}-\phi))}{2}.
\end{eqnarray}
The second order correction to the Jarskog CP invariant is given by
\begin{equation}
\triangle J \approx -\frac{\sin2\theta_{12}' \epsilon_{13}\epsilon_{23} \sin(\delta_{13}-2\phi)}{4\sqrt{2}} - \frac{\sin2\theta_{12}' \epsilon_{12}\epsilon_{23} \sin\phi}{4\sqrt{2}} - \frac{\cos2\theta_{12}' \epsilon_{12}\epsilon_{13} \sin(\delta_{13}-\phi)}{2}.
\end{equation}
The second order contributions to the $\nu_2.\nu_3$ unitarity triangle are given by
\begin{eqnarray}
\triangle U_{e2} U_{e3}^* \approx -\frac{c_{12}'(\epsilon^2_{12} - \epsilon^2_{13})}{2} + \frac{s_{12}' \epsilon_{23}(e^{-i\phi} \epsilon_{12} + e^{i\delta_{13}} \epsilon_{13})}{\sqrt{2}} + i c_{12}' \epsilon_{12} \epsilon_{13}\sin(\delta_{13}-\phi), \nonumber \\
\triangle U_{\mu2} U_{\mu3}^* \approx \frac{c_{12}'(\epsilon^2_{12}+2\epsilon^2_{23})}{2} - \frac{e^{-i\phi} s_{12}' \epsilon_{12}\epsilon_{23}}{\sqrt{2}} - i c_{12}'\epsilon_{12} \epsilon_{13}\sin(\delta_{13}-\phi), \nonumber  \\
\triangle U_{\tau2} U_{\tau3}^* \approx
-\frac{c_{12}'(\epsilon^2_{13}+2\epsilon^2_{23})}{2} - \frac{e^{i\delta_{13}} s_{12}' \epsilon_{13}\epsilon_{23}}{\sqrt{2}}. \ \ \ \ \ \ \ \ \ \ \ \ \ \ \ \ \
 \ \ \ \ \ \ \ \ \ \ \ \ \  
\end{eqnarray}
The second order contributions to the $\nu_e .\nu_{\mu}$ triangle are given by
\begin{eqnarray}
\triangle U_{\mu 1} U_{e1}^* \approx \frac{\sin2\theta_{12}'(4\epsilon^2_{12} + \epsilon^2_{13}+\epsilon^2_{23})}{4\sqrt{2}} + \frac{\sin2\theta_{12}' \epsilon_{12}\epsilon_{13}\cos(\delta_{13}-\phi)}{\sqrt{2}} + s_{12}'^{2} \epsilon_{12} \epsilon_{23}\cos\phi, \nonumber \ \ \\
\triangle U_{\mu 2} U_{e2}^* \approx -\frac{\sin2\theta_{12}'(4\epsilon^2_{12} + \epsilon^2_{13}+\epsilon^2_{23})}{4\sqrt{2}} - \frac{\sin2\theta_{12}' \epsilon_{12}\epsilon_{13}\cos(\delta_{13}-\phi)}{\sqrt{2}} + c_{12}'^{2} \epsilon_{12} \epsilon_{23}\cos\phi, \nonumber \\
\triangle U_{\mu 3} U_{e3}^* \approx - \epsilon_{12} \epsilon_{23}\cos\phi. \   \    
\ \ \ \ \ \ \ \ \ \ \ \ \ \ \ \ \ \ \ \ \ \ \ \ \ \ \ \ \ \ \ \ \ \ \ \ \ \ \ \ \ \ \ \ \ \ \ \ \ \ \ \ \ \ \ \ \ \ \ \ \ \ \ \ \ \ \ \ \ \ \ \ \
\end{eqnarray}


\begin{thebibliography}{99}
\bibitem{1} M. H. Ahn \textit{et al.}, [K2K Collaboration], \textit{Phys. Rev. Lett.} \textbf{90}, 041801 (2003), hep-ex/0212007; Y. Fukuda \textit{et al.}, [Super-Kamiokande Collaboration], \textit{Phys. Rev. Lett.} \textbf{81}, 1562 (1998), hep-ex/9807003; K. Eguchi \textit{et al.}, [KamLAND Collaboration], \textit{Phys. Rev. Lett.} \textbf{90}, 021802 (2003), hep-ex/0212021; T. Araki \textit{et al.}, [KamLAND Collaboration], \textit{Phys. Rev. Lett.} \textbf{94}, 081801 (2005), hep-ex/0406035; S. Abe \textit{et al.}, [KamLAND Collaboration], \textit{Phys. Rev. Lett.} \textbf{100}, 221803 (2008), arXiv:0801.4589 [hep-ex]; C. Arpesella \textit{et al.}, [Borexino Collaboration], \textit{Phys. Lett.} \textbf{B 658}, 101 (2008), arXiv:0708.2251 [astro-ph]; B. T. Cleveland \textit{et al.}, \textit{Astrophys.} \textbf{J 496}, 505 (1998); J. N. Abdurashitov \textit{et al.}, [SAGE Collaboration], \textit{J. Exp. Theor. Phys.} \textbf{95}, 181 (2002), astro-ph/0204245; W. Hampel \textit{et al.}, [GALLEX Collaboration], \textit{Phys. Lett.} \textbf{B 447}, 127 (1999). For recent reviews see e.g. S. F. King, \textit{Rept. Prog. Phys.} \textbf{67}, 107 (2004), hep-ph/0310204; R. N. Mohapatra, A. Y. Smirnov, \textit{Ann. Rev. Nucl. Part. Sci} \textbf{56}, 569 (2006), hep-ph/0603118; M. C. Gonzalez-Garcia, M. Maltoni, \textit{Phys. Rept.} \textbf{460}, 1-129 (2008), arXiv:0704.1800 [hep-ph]; T. Schwetz, M. Tortola and J. W. F. Valle, \textit{New. J. Phys.} \textbf{10}, 113011 (2008), arXiv:0808.2016 [hep-ph].
\bibitem{2} B. Pontecorvo,  \textit{Zh. Eksp. Teor. Fiz. (JTEP)} \textbf{33}, 549 (1957); \textit{ibid.} \textbf{34}, 247 (1958);  \textit{ibid.} \textbf{53}, 1717 (1967); Z. Maki, M. Nakagawa and S. Sakata, \textit{Prog. Theor. Phys.} \textbf{28} 870 (1962).
\bibitem{3} Particle Data Group (2010), K. Nakamura, \textit{et. al.}, \textit{J. Phys.} \textbf{G}: \textit{Nucl. Part. Phys.} \textbf{37}, 075021.
\bibitem{4} P. F. Harrison, D. H. Perkins and W. G. Scott, \textit{Phys. Lett.} \textbf{B 530}, 167 (2002), hep-ph/0202074; P. F. Harrison and W. G. Scott, \textit{Phys. Lett.} \textbf{B 535}, 163 (2002), hep-ph/0203209; Zhi-zhong Xing, \textit{Phys. Lett.} \textbf{B 533}, 85 (2002), hep-ph/0204049.
\bibitem{5} For recent reviews and list of references, see e.g. G. Altarelli and F. Feruglio, \textit{Rev. Mod. Phys.} \textbf{82}, 2701 (2010), arXiv:1002.0211 [hep-ph]; H. Ishimori, T. Kobayashi, H. Ohki, H. Okada, Y. Shimizu and M. Tanimoto, \textit{Prog. Theor. Phys. Suppl.} \textbf{183}, 1-163 (2010), arXiv:1003.3552 [hep-ph].
\bibitem{6} A. Datta, F. S. Ling and P. Ramond, \textit{Nucl. Phys.} \textbf{B 671}, 383 (2003), hep-ph/0306002; Y. Kajiyama, M. Raidal and A. Strumia, \textit{Phys. Rev.} \textbf{D 76}, 117301 (2007), arXiv:0705.4559 [hep-ph]; L. L. Everett, A. J. Stuart, \textit{Phys. Rev.} \textbf{D 79}, 085005 (2009), arXiv:0812.1057 [hep-ph]; F. Feruglio, A. Paris, \textit{JHEP} \textbf{1103}, 101 (2011), arXiv:1101.0393 [hep-ph].
\bibitem{7} W. Rodejohann, \textit{Phys. Lett.} \textbf{B 671}, 267 (2009), arXiv:0810.5239 [hep-ph]; A. Adulpravitchai, A. Blum and W. Rodejohann, \textit{New J. Phys.} \textbf{11}, 063026 (2009), arXiv:0903.0531 [hep-ph].
\bibitem{8}C. H. Albright, A. Dueck and W. Rodejohann, \textit{Eur. Phys. J.} \textbf{C 70}, 1009 (2010), arXiv:1004.2798 [hep-ph].
\bibitem{9} F. Vissani, hep-ph/9708483; V. D. Barger, S. Pakvasa, T. J. Weiler and K. Whisnant, \textit{Phys. Lett.} \textbf{B 437}, 107 (1998), hep-ph/9806387; A. J. Baltz, A. S. Goldhaber and M. Goldhaber, \textit{Phys. Rev. Lett.} \textbf{81}, 5730 (1998), hep-ph/9806540.
\bibitem{10}C. I. Low, R. R. Volkas, \textit{Phys. Rev.} \textbf{D 68}, 033007 (2003) hep-ph/0305243; C. S. Lam \textit{Phys. Rev.} \textbf{D 74}, 113004 (2006), hep-ph/0611017; I. de Medeiros Varzielas, R. Gonz\'{a}lez Felipe, H. Ser\^{o}dio, \textit{Phys. Rev.} \textbf{D 83}, 033007 (2011), arXiv:1101.0602 [hep-ph].
\bibitem{11} G. L. Fogli, E. Lisi, A. Marrone, A. Palazzo, A. M. Rotunno, \textit{Phys. Rev. Lett.} \textbf{101}, 141801 (2008), arXiv:0806.2649 [hep-ph]; M. C. Gonzalez-Garcia, M. Maltoni, J. Salvodo, \textit{JHEP} \textbf{04}, 056 (2010), arXiv:1001.4524 [hep-ph].
\bibitem{12} The T2K collaboration, K. Abe et al., \textit{Phys. Rev. Lett.} \textbf{107}, 041801 (2011), arXiv:1106.2822 [hep-ex].
\bibitem{13} M. Jezabek, Y. Sumino, \textit{Phys. Lett.} \textbf{B 457}, 139 (1999), hep-ph/9904382; Z. Z. Xing, \textit{Phys. Rev.} \textbf{D 64}, 093013 (2001), hep-ph/0102304; C. Giunti, M. Tanimoto, \textit{Phys. Rev.} \textbf{D 66}, 053013 (2002), hep-ph/0207096; \textit{ibid.} \textbf{66}, 113006 (2002), hep-ph/0209169; W. Rodejohann, \textit{Phys. Rev.} \textbf{D 69}, 033005 (2004), hep-ph/0309249; N. Li, B. Q. Ma, \textit{Phys. Lett.} \textbf{B 600}, 248 (2004), hep-ph/0408235; G. Altarelli, F. Feruglio and I. Masina, \textit{Nucl. Phys.} \textbf{B 689}, 157 (2004) hep-ph/0402155; A. Romanino, \textit{Phys. Rev.} \textbf{D 70}, 013003 (2004), hep-ph/0402258.
\bibitem{14} P. H. Frampton, S. T. Petcov and W. Rodejohann, \textit{Nucl. Phys.} \textbf{B 687}, 31 (2004), hep-ph/0401206.
\bibitem{15} Z. Z. Xing, \textit{Phys. Lett.} \textbf{B 533}, 85 (2002), hep-ph/0204049; R. N. Mohapatra, W. Rodejohann, \textit{Phys. Rev.} \textbf{D 72}, 053001 (2005), hep-ph/0507312; S. F. King, \textit{JHEP} \textbf{0508}, 105 (2005), hep-ph/0506297; I. Masina, \textit{Phys. Lett.} \textbf{B 633}, 134 (2006), hep-ph/0508031; S. Antusch, S. F. King, \textit{Phys. Lett.} \textbf{B 631}, 42 (2005), hep-ph/0508044; S. Antusch, P. Huber, S. F. King and T. Schwetz, \textit{JHEP} \textbf{0704}, 060 (2007) hep-ph/0702286; S. Boudjemaa, S. F. King, \textit{Phys. Rev.} \textbf{D 79}, 033001 (2009), arXiv:0808.2782 [hep-ph];Y. Shimizu, R. Takahashi, \textit{EPL} \textbf{93}, 61001 (2011), arXiv:1009.5504 [hep-ph]; D. Meloni, F. Plentinger, W. Winter \textit{Phys. Lett.} \textbf{B 699}, 354 (2011), arXiv:1012.1618 [hep-ph]; A. S. Joshipura, K. M. Patel, arXiv:1105.5943 [hep-ph]; Ya-juan Zheng, Bo-Qiang Ma, arXiv:1106.4040 [hep-ph]; Xiao-Gang He, A. Zee, arXiv:1106.4359 [hep-ph]; Y. H. Ahn, Hai-Yang Cheng, S. Oh, arXiv:1107.4549 [hep-ph]. 
\bibitem{16} F. Plentinger, W. Rodejohann, \textit{Phys. Lett.} \textbf{B 625}, 264 (2005), hep-ph/0507143.
\bibitem{17} K. A. Hochmuth. S. T. Petcov, W. Rodejohann, \textit{Phys. Lett.} \textbf{B 654}, 177 (2007), arXiv:0706.2975 [hep-ph].
\bibitem{18}C. H. Albright, W. Rodejohann \textit{Eir. Phys. J.} \textbf{C 62}, 599 (2009), arXiv:0812.0436 [hep-ph]; C. H. Albright, A. Dueck, W. Rodejohann \textit{Eir. Phys. J.} \textbf{C 70}, 1099 (2010), arXiv:1004.2798 [hep-ph].
\bibitem{19} S. Pascoli, S. T. Petcov and W. Rodejohann, \textit{Phys. Rev.} \textbf{D 68}, 093007 (2003), hep-ph/0302054.
\bibitem{20} G. L. Fogli, E. Lisi, A. Marrone, A. Palazzo, A. M. Rotunno, arXiv:1106.6028 [hep-ph].
\bibitem{21} L. Whitehead [MINOS Collaboration], $http://www-numi.fnal.gov/pr_-plots$
\bibitem{22} C. Jarlskog, \textit{Phys. Rev. Lett.} \textbf{55}, 1039 (1985).
\bibitem{23} S. F. King, \textit{Phys. Lett.} \textbf{B 659}, 244 (2008), arXiv:0710.0530 [hep-ph].

\end{thebibliography}
\end{document}